\documentclass[prb,floatfix,superscriptaddress,showpacs,twocolumn,a4paper]{revtex4}
\usepackage{amssymb,amsmath}
\usepackage{graphicx}
\usepackage{color}
\usepackage{bbm}

\usepackage{mathrsfs} 

\newcommand{\ie}{i.e.\ }
\newcommand{\eg}{e.g.\ }
\def\se{self-energy}

\def\bs{band-structure}
\def\uc{unit cell}

\def\etal{{\it et al.}}

\def\t2g{t\ensuremath{_{2g}}}
\def\egs{e\ensuremath{_g^\sigma}}
\def\egp{e\ensuremath{_g^\pi}}
\def\a1g{a\ensuremath{_{1g}}}

\def\vo2{VO\ensuremath{_2}}
\def\v2o3{V\ensuremath{_2}O\ensuremath{_3}}

\newcommand{\bra}[1]{\langle #1|}
\newcommand{\ket}[1]{|#1\rangle}
\newcommand{\braket}[2]{\langle #1|#2\rangle}

\newcommand{\eref}[1]{Eq.~(\ref{#1})}
\newcommand{\fref}[1]{Fig.~\ref{#1}}
\newcommand{\sref}[1]{Section~\ref{#1}}

\newcommand{\vek}[1]{%
        \hbox{\textbf #1}}

\newcommand{\mop}{%
        \mathbf}
\newcommand{\op}[1]{%
        \hbox{\textbf #1}}
\newcommand{\pr}{%
        ^\prime}

\newcommand{\ppr}{%
        ^{\prime\prime}}

\newcommand{\com}[1]{%
        \left[#1\right]}

\newcommand{\svek}{%
        \mathbf}
\newcommand{\im}{%
        {\imath}}     
\newcommand{\herm}{%
        {\cal H}} 
\newcommand{\ddr}{%
        \hbox{d$^3$r }}
        
\newcommand{\dif}{%
        \hbox{d}}           

\newcommand{\To}{%
        {\cal T }}             
        
\newcommand{\cc}{%
        c^\dag}    
\newcommand{\ca}{%
        c^{\phantom{\dag}}}

\newcommand{\tr}{%
        \hbox{ tr}}

\newcommand{\inp}{%
        \im 0^+} 

\newcommand{\fermi}[1]{%
        \hbox{f($#1$)}}

\begin{document}

\title{Optical properties of correlated materials -- \\ Generalized Peierls approach and its application to VO$_2$}

\author{Jan M. Tomczak}
\affiliation{Research Institute for Computational Sciences, AIST, Tsukuba, 305-8568 Japan}
\affiliation{Japan Science and Technology Agency, CREST}
\author{Silke Biermann}
\affiliation{Centre de Physique Th{\'e}orique, Ecole polytechnique, CNRS, 91128 Palaiseau, France}
\affiliation{Japan Science and Technology Agency, CREST}
\date{\today }

\begin{abstract}
The aim of the present paper is to present a versatile scheme for the computation of optical properties of solids, with particular emphasis on realistic many-body calculations for correlated materials. 
Geared at the use with localized basis sets, we extend the commonly
known lattice ``Peierls substitution'' approach to the case of
multi-atomic unit cells. We show in how far this generalization can be deployed as an approximation to the full Fermi velocity matrix elements that enter the continuum description of the response  of a solid to incident light. 
We further devise an upfolding scheme to incorporate optical transitions, that involve high energy orbitals that had been downfolded in the underlying many-body calculation of the electronic structure.
As an application of the scheme, we present results on a material of longstanding interest, vanadium dioxide, VO$_2$.
Using dynamical mean-field data of both, the metallic and the insulating
phase, we calculate the
corresponding optical conductivities, elucidate optical transitions and find good agreement with experimental results.
\end{abstract}

\maketitle

\section{Introduction}

Correlated matter is characterized by an enormous sensitivity with respect to changes of external parameters. It is the merit of this responsiveness that a remarkable richness of properties emerges in these systems. Correlation effects seem, for instance, to be a vital issue to outstanding phenomena such as high temperature superconductivity and colossal magnetoresistance.
In the latter case, the possibility of tuning the fundamental behavior
of a material by an external field will undoubtedly lead to yet improved data storage devices. 
A better understanding of the various effects of strong correlations is
thus a highly desirable goal of condensed matter physics, both from
the theoretical and the technological point of view.

On the experimental side, 
numerous techniques have been devised for and applied to the study of correlated materials of ever growing complexity. Optical spectroscopy, which is the subject of this paper, is, in a way, the most natural among them~: Optical detectors are sampling the response to incident light, as do our eyes,
albeit accessing frequencies, and thus phenomena, that are beyond our vision.
The technique is particularly suited for tracking the evolution of the
system under changes of, for instance, temperature or pressure.
 This is owing to a generally high precision, and the fact that, contrary to e.g.\ photoemission spectroscopy or x-ray experiments, results are obtained in absolute values. Especially, the existence of sum-rules allows for a quantitative assessment of transfers of spectral weight upon changes of the system properties.
Therewith optical spectroscopy is particularly adapted to the study of
correlated materials~\cite{PhysRevB.73.165116,PCCO,Qazilbash12142007,Baldassarre_v2o3,pnictide_optic}.

On the theory side,
while weakly correlated materials are well described within density functional theory (DFT)\cite{RevModPhys.71.1253}, e.g.\ in the local density approximation (LDA)\cite{kohn2}, and moderate correlation effects are captured by perturbative approaches, such as Hedin's GW approximation\cite{hedin}, it was 
the advent of dynamical mean-field theory (DMFT)(for reviews see e.g.\ \cite{bible,vollkot}), and its realistic extension, LDA+DMFT (for reviews see \cite{held_psik,biermann_ldadmft}),
which allowed for the description and understanding of several metal-insulator transitions that are derived from the Mott-Hubbard or related mechanisms. 
Though our discussion of optical properties within realistic many-body approaches is quite general, and applicable to other techniques, we present results on \vo2\ that are based on LDA+DMFT calculations.

The paper is organized as follows~: 
After having now expanded on the potency of optical spectroscopy and the motivation for more theoretical efforts, we will
in the remainder of Section I  briefly review experimental and theoretical knowledge about  vanadium dioxide.
In Section II we develop our formalism for the optical conductivity within realistic calculations. 
Section III is devoted to a detailed discussion on Fermi velocities. This part contains our major innovations.
Readers less interested in technical details are welcome to jump directly to Section IV, which presents our theoretical optics spectra for vanadium dioxide in both, the metallic and the insulating phase.

\subsection{Vanadium dioxide -- The material}
\subsubsection{Basic electronic structure}

At its metal-insulator
transition~\cite{PhysRevLett.3.34} (T$_c$=340~K), \vo2\ transforms from a metallic high temperature phase of
rutile structure into an insulating monoclinic (M1) phase,
in which the vanadium atoms pair up to form tilted dimers along the
c-axis. This M1 phase was found to be non-magnetic~\cite{pouget_review}.

Over the decades several scenarios were evoked to explain the
metal-insulator transition.
This is not the emphasis of the current paper, and we shall only
briefly summarize the basic electronic structure. 

\vo2\ has a vanadium 3d$^1$ configuration and the crystal field splits the
3d-manifold into \t2g\ and empty \egs\ components. The
former are further split into \egp\ and \a1g\ orbitals.  In the rutile
phase these orbitals overlap, accounting for the metallic character.
In the M1 phase, the \a1g\ split into bonding/anti-bonding orbitals,
due to the aforementioned vanadium dimerization.

The Goodenough scenario~\cite{PhysRev.117.1442, goodenough_vo2} of the
insulator advocates the structural effect of the
unit-cell doubling due to the dimer formation as the main origin of
the gap-formation, and thus attributes the insulating behavior to a
Peierls transition\cite{peierls_book}. Zylbersztejn and Mott~\cite{PhysRevB.11.4383} on the other hand stressed the
importance of local Coulomb interactions, and thought the transition
to be of, what we call today, the Mott-Hubbard type.
Experiments~\cite{PhysRevLett.35.873,PhysRevB.10.1801,koethe:116402,haverkort:196404,eguchi_vo2,qazilbash:075107} were interpreted to support one or the other of the two scenarios.
Important to note is that neither of the two phases are well-described
within standard band-theory approaches. In the rutile phase these
miss bandwidth-narrowing and satellite features, as seen \eg in
photoemission~\cite{koethe:116402}, and the bad metal conductivity
seen in transport measurements~\cite{ladd_vo2}. In the M1 phase, the
problem is even more fundamental, since band-theory fails to produce
an insulating behavior\cite{PhysRevLett.72.3389,eyert_vo2,korotin_vo2}.
For reviews see \eg\cite{imada, me_phd}.
As we will detail later, the LDA+DMFT approach succeeds in describing the
experimental findings of both, the metallic~\cite{0295-5075-69-6-984,liebsch:085109, biermann:026404} and the insulating~\cite{biermann:026404} phase.
As a matter of fact the current
calculations of optical properties rely on our previous LDA+DMFT work~\cite{biermann:026404,tomczak_vo2_proc,me_vo2}
which in particular extended on the interpretation of the nature of the insulating M1
phase. Compatibility with experimental results on the optical
conductivity strengthens this picture. 
For the insulating phase, also GW type of calculations~\cite{PhysRevB.60.15699,me_phd,sakuma:075106,gatti:266402} as well as LDA+U 
based approaches~\cite{korotin_vo2,Eckern_vo2} open the charge gap.
Further, VO$_2$ has also been studied within cluster based methods, see e.g.~\cite{tanaka_vo2,Mossanek2005189,mossanek:125112}.

\subsubsection{Insights by optical measurements}

Optical measurements on \vo2\ were first performed by Barker \etal
\cite{PhysRevLett.17.1286}, and Verleur
\etal\cite{PhysRev.172.788}. By probing different orientations of
single crystal samples, they
evidenced an anisotropy in the optical response of the M1 insulator.
More precisely, the conductivity depends on whether
the electric field is parallel or perpendicular to the
crystallographic rutile c-axis when going below the transition temperature.
This is to be expected from the changes in the
crystal-structure and the unit-cell doubling along the c-axis.
This anisotropy was confirmed by ultraviolet reflectance
measurements~\cite{PhysRevB.41.4993} and
x-ray experiments\cite{PhysRevB.43.7263,koethe:116402} (see also \cite{PhysRevB.20.1546}).
Ladd \etal~\cite{ladd_vo2} performed experiments under pressure, and noticed that c-axis stress reduces the transition temperature considerably more than is the case for hydrostatic pressure. 
Okazaki \etal~\cite{PhysRevB.73.165116} studied reflectance spectra of thin films with an orientation of the electric field perpendicular to the rutile c-axis as a function of temperature, and found indications for electron-phonon coupling.
Recent studies by Qazilbash \etal~\cite{qazilbash:205118} (see also \cite{qazilbash:115121,qazilbash:115121}) on polycrystalline films
with preferential $[010]$ orientation~\cite{chae:C12} confirmed the bad metal
behavior of rutile
\vo2\ evidenced in transport experiments~\cite{ladd_vo2}. Indeed, rutile \vo2\ is found to violate~\cite{qazilbash:205118} the Ioffe-Regel-Mott limit for
resistivity saturation~\cite{RevModPhys.75.1085},
\ie the electron mean free path is comparable to, or smaller than the lattice spacing
and Boltzmann transport theory breaks down.
As we shall see below, pronounced differences
in the optical response are found between the individual experiments.

Theoretically, the optical response of M1 \vo2\ was investigated by means
of a self-consistent model GW calculation by Continenza \etal\cite{PhysRevB.60.15699},
which was found to improve on LDA results for the dielectric function,
when comparing with experiments~\cite{PhysRevB.5.3138}. Also, a clear
polarization dependence was evidenced. 

Further, the dielectric response of both the metallic and the
insulating phase were calculated within
LDA by  Mossanek and Abbate\cite{0953-8984-19-34-346225}.
In the metallic phase, peak positions and the polarization dependence
were qualitatively captured. The issue of the bad metallic behavior
was not addressed, which is natural since it lies 
beyond band theory.
As to the insulating M1 phase, a rigid shift was introduced to the LDA
band-structure, such as to ``artificially'' produce a gap. This procedure,
again, resulted in qualitative agreement with experiment.
However, we believe that the electronic structure is characterized by
an enhanced \a1g\ bonding/anti-bonding splitting~\cite{me_vo2}, which is
not reproduced by an orbital-independent shift. 
An orbital-dependent one-particle potential, on the other hand, actually does 
correctly capture spectral properties to a surprising degree~\cite{tomczak_vo2_proc,me_vo2}.
We will come back to this in Section III.

\section{Optical conductivity in realistic calculations}

\subsection{Optical conductivity from DMFT calculations}

Within the field of strongly correlated electrons, calculations of the
optical conductivity within the DMFT framework were first performed by
Jarrell \etal\cite{PhysRevB.47.3553}, and Pruschke \etal\cite{ PhysRevB.51.11704}
for the case of the Hubbard model.
Rozenberg
\etal\cite{PhysRevLett.75.105,PhysRevB.54.8452} studied the phenomenology of the different optical
responses of the Hubbard model throughout its phase diagram 
in conjunction with experiments on V$_2$O$_3$.
In the realistic LDA+DMFT context, optical conductivity calculations were first
performed by Bl{\"u}mer\cite{bluemer, blumer-2003} for the case of
degenerate orbitals.
A more general approach, was developed in~\cite{palsson} for the study
of transport properties.
Further, recent LDA+DMFT works that use simplified approaches to the Fermi velocities can be found in~\cite{1367-2630-7-1-188,Baldassarre_v2o3}.
Our work goes along the lines of the mentioned approaches. We will however use a
full Hamiltonian formulation, therewith allowing for the general case of
non-degenerate orbitals, and we extend the intervening Fermi
velocities to multi-atomic \uc s, which becomes crucial in calculations for realistic compounds.

Alternative techniques were presented by Perlov \etal~\cite{PhysRevB.68.245112} in the  
Korringa-Kohn-Rostoker (KKR) context, and by
Oudovenko \etal\cite{oudovenko:125112}.
The idea in the latter work is to diagonalize the interacting system, which allows for
the analytical performing of some occurring integrals due to the
``non-interacting'' form of the Green's function. Owing to the
frequency-dependence of the \se, however, the
diagonalization has to be performed for each momentum and frequency
separately, so the procedure may become
numerically expensive.

First accounts of the presented optics scheme have been given in Ref. \cite{tomczak_v2o3_proc} for V$_2$O$_3$ (based on the electronic structure of Ref.~\cite{poter_v2o3}), while
applications can be found in Ref.~\cite{optic_epl,me_psik}.

\subsection{The optical conductivity}

The optical conductivity tensor $\sigma^{\alpha\beta}(\vek{q},\omega)$ is defined as the linear response that
relates the total electric field in the solid to the charge current
density \cite{mahan}~:
\begin{eqnarray}\label{defoptcond}
    \langle j^\alpha(\vek{r},t)\rangle=\sum_\beta\sigma ^
    {\alpha\beta}(\vek{q},\omega)E ^\beta(\vek{r},t)
\end{eqnarray}
here $\alpha,\beta$ denote cartesian coordinates and
$\langle\cdot\rangle$ indicates the quantum mechanical expectation
value.

In the following we will derive an expression for the long wavelength
limit ($\vek{q}=0$) of the real part of the conductivity tensor, 
which we shall refer to as the optical conductivity $\Re\sigma
^{\alpha\beta}(\omega)$. The starting
point of the derivation is the fundamental Hamiltonian of the system,
$\op{H}=\op{H}_0^A+\op{H}_{int}$, with $\op{H}_0^A$ being the
one-particle part, with the coupling to the (classical) light field via its vector potential $\vek{A}(\vek{r},t)$~:
\begin{equation}\label{H0}
\sum_\sigma\int\ddr\Psi^\dag_\sigma(\vek{r},t)\left[\frac{1}{2m}\bigl(\im\hbar\nabla +\frac{e}{c}\vek{A}(\vek{r},t)\bigr)^2 + V(\vek{r})   \right]\Psi^{\phantom{\dag}}_\sigma(\vek{r},t)
\end{equation}
where we have chosen the Coulomb gauge, $\nabla \cdot\vek{A}=0$.  
$V(\vek{r})$ is any one-particle potential. In practice, it will e.g.\ be the effective Kohn-Sham potential of density functional theory within the LDA. We emphasize that 
this notation in terms of the field operators, $\Psi$, is still basis free.
$\op{H}_{int}$ contains, in our case, electron-electron
 interactions, involving only two-body terms.
 Then, it commutes with the
charge density and is trivially gauge invariant.
\footnote{The restriction to two body terms is less severe than
  the assumption of Hubbard-Hund type interactions
 restricted to density-density terms only, which is necessary when
 starting from a lattice formulation, as in \eref{latticeH}.}

Gauge invariance of the full Hamiltonian leads to charge conservation
 and, via the continuity equation $  e
 \partial_t\rho=-\nabla\cdot \vek{j}$,  we obtain the expression for
 the charge current density $j^\alpha(\vek{r},t)$
\begin{eqnarray}
\label{current}
-\herm\left\{\im\frac{e\hbar}{m} \sum_\sigma\Psi_\sigma^\dag(\vek{r},t) \nabla_\alpha
  \Psi^{\phantom{\dag}}_\sigma(\vek{r},t)\right\}+\frac{e^2}{mc}A ^\alpha(\vek{r},t)\rho(\vek{r},t)\nonumber\\
\end{eqnarray}
where $\herm$ denotes the hermitian part, and we assumed the vector potential to be directed along the $\alpha$-direction.
In the Coulomb gauge, the second
term in \eref{current} is the diamagnetic current, which we will drop
hereafter, since its contribution to the conductivity is purely
imaginary. The first term is called the paramagnetic current. For a
discussion on their physical interpretation see \eg \cite{coleman}.
The current expectation value is, within the Kubo linear response formalism, linked to
the current-current correlation function. For the optical
conductivity, we then find
\begin{eqnarray}
\Re\sigma ^{\alpha\beta}(\omega)&=&-\frac{\Im\chi ^{\alpha\beta}(\omega+\inp)}{\omega}
\end{eqnarray}
where $\chi(\omega)$ is the longwavelength-limit ($\vek{q}=0$) of 
\begin{eqnarray}\label{chi}
        \chi ^{\alpha\beta}(\vek{q},\tau)&=&-\frac{1}{\hbar V}\left\langle\To
        j^\alpha(-\vek{q},\tau),j^\beta(\vek{q},0)
        \right\rangle
\end{eqnarray}
which we have written in the imaginary time Matsubara formalism.
So far, all quantities live in the spatial continuum. 
As announced, we shall make the connection with many-body techniques that work in localized basis sets.
At this point, however, we shall first develop the field operators in a Bloch-like basis, $\Psi
(\vek{r},\tau)=
\sum_{\svek{k}L\sigma} \braket{\vek{k}L}{r}
\ca_{\svek{k}L\sigma}(\tau)$. Here, $L=(n,l,m,\gamma)$ denotes orbital $(n,l,m)$ of atom
$\gamma$ within the \uc. $\cc_{\svek{k}L\sigma}$, $\ca_{\svek{k}L\sigma}$ are the usual (discrete) creation and annihilation operators. 
The momentum  sum runs
over the first Brillouin zone. Later on, we will switch to the
Wannier-like, 
real space basis
$\ket{\vek{R}L}= 
\sum_\svek{k}\exp{(\im\svek{k}\svek{R})}\ket{\svek{k}L}$, to which the aforementioned notion of localization will apply.
Here, $\vek{R}$ labels the unit-cell, which only in case of a
one-atomic basis is equivalent to the atomic position. 
 This distinction
will prove important later on.
Taking the limit of long wavelengths, which in this context is the familiar
dipole approximation, we find for the paramagnetic current
\begin{eqnarray}
{j} ^{\alpha}(\vek{q}=0,\tau)=e
\sum_{\svek{k},LL\pr,\sigma}v_{\svek{k},\alpha}^{L\pr L}\cc_{\svek{k}\pr L\pr\sigma}(\tau)\ca_{\svek{k}L\sigma}(\tau)
\end{eqnarray}
with the so-called Fermi velocity, or dipole matrix element
\begin{eqnarray}\label{matel}
v_{\svek{k},\alpha}^{L\pr L}&=&
\frac{1}{m}\herm\bra{\svek{k} L\pr}\mathcal{P}_\alpha\ket{\svek{k}L}
\end{eqnarray}
where $\mathcal{P}_\alpha$ is the $\alpha$-component of the momentum operator.
In the evaluation of $\vek{q}=0$ correlation functions in infinite dimensions, thus in a dynamical mean-field spirit,
vertex corrections are absent  in the one orbital
case\cite{PhysRevLett.64.1990,bible}. In other words, electron-hole interactions effectively vanish, and
the two-particle quantity, \eref{chi}, can be decoupled into the product of two one-particle Green's functions. 
In the following, we shall neglect vertex corrections, though the above
statement is not valid for the cluster or multi orbital case.
Indeed, one can show that only the elements of the Fermi velocity that
are diagonal in orbital space have the odd parity with respect to
momentum that is required for the vanishing of vertex correction in $d=\infty$.
Instead of advancing towards a more stringent two-particle formulation in the model case, it is our objective to strive after a formalism that accounts for the complexity of realistic state-of-the-art electronic structure techniques, such as LDA+DMFT.
With this simplification, the derivation, after continuing
to real frequencies ($\im\omega_n\rightarrow\omega+\im0^+$), yields~:
\begin{eqnarray}\label{oc}
\Re\sigma ^{\alpha\beta} (\omega)&=&\frac{2\pi e ^2\hbar}{V}\sum_{\svek{k}}\int\dif\omega\pr\;\frac{\fermi{\omega\pr}-\fermi{\omega\pr+\omega}}{\omega}\nonumber\\
&\times&\tr\biggl\{ A(\vek{k},\omega\pr+\omega) v_\alpha(\vek{k}) A(\vek{k},\omega\pr)  v_\beta(\vek{k}) \biggr\}\label{oc_para2}
\end{eqnarray}
Where $\tr$ stands for the trace over orbitals and $
A(\vek{k},\omega)=-1/\pi\,\Im\left[\omega+\mu-\op{H}_0(\svek{k})-\Sigma(\omega)\right]^{-1}$ and $v_\alpha(\vek{k})$ are the orbital matrices of
the momentum-resolved spectral functions and the Fermi velocities,
respectively. $\op{H}_0$ denotes the one-particle Hamiltonian in the
absence of the external field.
Hence, in total, 
the conductivity acquires the well-known form of a frequency convolution of momentum
resolved spectral functions, with the Fermi velocities modulating the amplitudes of the spectral weight.
Due to the fact, mentioned earlier, that the interaction part of the Hamiltonian
commutes with the charge density, the Fermi velocities are those of
the non-interacting problem $\op{H}_0$. The many-body physics only
enters the
spectral functions via the evaluation of the expectation value 
in the correlation function of \eref{chi}.

\section{Fermi velocities}

The Fermi velocity matrix element, \eref{matel}, is readily evaluated when e.g.\ working within a plane wave basis set. Yet, many-body techniques, such as DMFT and its realistic extensions, that are geared at improving on local interactions in the spirit of the Hubbard model, necessitate the use of localized orbital sets, 
 \eg muffin tin derived orbitals, $L/N$MTO\cite{lmto,nmto}, or other
 Wannier functions\cite{lechermann:125120}. While a computation of the
 full matrix element, \eref{matel}, is in principle still possible within these basis sets, it becomes rather tedious from the practical point of view.
Therefore we have devised a handy approximation, explicitly geared at the use with localized orbitals, which allows for a reliable calculation of optical properties at a rather low computational cost~\cite{optic_epl,me_psik}.

\subsection{Lattice formulation~: \\ The Peierls substitution and its generalization}

In the above, we coupled the light field to the electronic degrees of freedom of the solid via the standard minimum coupling, and developed the continuous field operators into a basis, which led to the given Fermi velocities of \eref{matel}.
A different approach is to instead develop first the Hamiltonian in this basis, and to couple the vector potential directly to the site, or lattice operators in a way that verifies gauge invariance.
Consider the Hubbard model
\begin{eqnarray}\label{latticeH}
        \op{H}&=&-\sum_{ij,LL\pr,\sigma}t_{ij}^{L\pr L}\cc_{iL\pr\sigma}\ca_{jL\sigma} 
        +\op{H}_{int}
\end{eqnarray}
 Then, the philosophy of the ``Peierls substitution''~\cite{peierls,millis_review} approach is to add the following phase-factors to the lattice operators\cite{millis_review}~: $\cc_{iL\sigma}\rightarrow\cc_{iL\sigma}\exp\left(\im\frac{e}{c\hbar}\int^{\svek{R}_{iL}}d\svek{r}\, \svek{A}(\svek{r},t)\right)$. Here, $\vek{R}_{iL}$ denotes
the atomic positions. These can be separated into
\begin{equation}\label{rpos}
\vek{R}_{iL}=\vek{R}_i+\mop{\rho}_\gamma
\end{equation} 
where the former indexes the
\uc\ $i$ and the latter the atom $\gamma$ within the cell, in the case of a multi-atomic \uc.
This seemingly trivial statement will lead to important terms in the Fermi velocities, which to our knowledge have so far not been considered.
Equally, in a multi-atomic environment, the lattice position operator $\mathcal{R}$ can be defined as
\begin{eqnarray}\label{Rop}
\mathcal{R}=\sum_{iL\sigma}\mop{R}_{iL}\cc_{iL\sigma}\ca_{iL\sigma}  
\end{eqnarray}
When now supposing the interactions in
\eref{latticeH} to be only of density-density type {\it in the lattice operators} (cf. footnote above), then the above phases
only appear in the kinetic part of the Hamiltonian. When additionally
assuming a slowly varying vector potential such as to approximate the
integral in the exponent, the Peierls approach can also be seen as a
substitution for the hopping amplitudes in the above lattice Hamiltonian~:
 $ t^{L\pr L}_{ij}\rightarrow t^{L\pr L}_{ij}\exp\left(\im \frac{e}{\hbar c} \svek{A}(t)\,(\svek{R}_{iL\pr}-\svek{R}_{jL})\right)
$.
We further remark that evidently the vector potential only
couples to non-local hopping elements, \ie within this approach
intra-atomic transitions, $\bigl(i,L=(n,l,m,\gamma)\bigr)\rightarrow \bigl( i,\widetilde{L}=(n\pr,l\pr,m\pr,\gamma)\bigr)$ in the above notation, are absent.
From the thus defined substitution, we can compute the current either by means of the continuity equation, or by a functional derivative of the Hamiltonian with respect to the vector potential. One finds
\begin{eqnarray}
        \vek{j}^\alpha&=&e\sum_{LL\pr,\svek{k},\sigma}v^{L\pr L}_{\svek{k}\, \alpha}\,\cc_{\svek{k} L\pr\sigma}\ca_{\svek{k}L\sigma} 
\end{eqnarray}
with the velocity
\begin{eqnarray}\label{vkelement}
        v^{L\pr L}_{\svek{k}\,
          \alpha}&=&\frac{\im}{\hbar}\sum_{ij}t_{ij}^{L\pr
          L}(R^\alpha_{iL\pr}-R^\alpha_{jL}) e^{-\im\svek{k}(\svek{R}_{i}-\svek{R}_{j})}
\end{eqnarray}
Using the separation \eref{rpos}, and introducing the usual Hamiltonian element 
$\op{H}_\svek{k}^{L\pr L}=-\sum_{ij}t_{ij}^{L\pr L}e^{-\im\svek{k}(\svek{R}_{i}-\svek{R}_{j})}$,
we find a generalized Peierls expression~:
\begin{eqnarray}
        \label{vkpeierls}
        v^{L\pr L}_{\svek{k}\,\alpha}=\frac{1}{\hbar} \biggl(
        \partial_{k_\alpha}\op{H}^{L\pr L}_{\svek{k}} -\im
        (\rho_{L\pr}^\alpha-\rho_L^\alpha)\op{H}^{L\pr L}_{\svek{k}} \biggr)
\end{eqnarray}
The first term is the familiar Fermi velocity, given by the momentum derivative of the
Hamiltonian. It contains hopping processes
that take place between {\it different \uc s} $i,j$. While absent
in the one-atomic case, $\gamma=\gamma\pr$, the second term, which to our knowledge is new, becomes crucial, once calculations
of realistic materials are performed. It accounts for hopping
amplitudes between {\it different atoms} $\gamma\pr,\gamma$ within the {\it same} \uc.
Indeed when considering \eg a simple cubic one-atomic system, first in
its primitive unit-cell, and then in a non-primitive \uc\ that is
doubled in the direction along which the momentum derivative is taken.
Then it is the
second term in \eref{vkpeierls} that assures that the 
optical conductivities of the two equivalent descriptions are the same.
 When using the derivative term only, the
optical conductivity comes out wrong, even
qualitatively.

The above formula is very handy, since the only intervening matrix element is that of the Hamiltonian and its momentum derivative. No other
integrals involving the LDA wave functions, which are cumbersome to handle in the chosen basis, occur in this case.
We note that the above expression is hermitian. Yet, in general, it has no well defined parity with respect to the momentum $\vek{k}$, even when assuming inversion symmetry of the Hamiltonian. Only the elements that are diagonal in the atomic $\gamma$-indices  have the required odd parity that leads to the cancellation of vertex corrections in the limit of infinite coordination. 

\subsection{Continuum formulation~: \\ Assessing the Peierls substitution} 

In the preceding section, an expression for the Fermi velocity was deduced from the {\it lattice} formulation of the solid.
This has to be contrasted to the proper matrix element, \eref{matel}, that originates from the {\it continuum} description.
Now, a valid question is whether, and under which circumstances, the generalized Peierls velocity can be employed as a
reliable approximation to the true dipole matrix element.
In the appendix we show that the latter can actually quite naturally be split into the generalized Peierls expression and
a correction term that recovers the full matrix element.
The impact of this supplementary term decreases with an increasing localization of the basis functions.
Indeed, in the limit of strongly localized orbitals, the only missing terms are
atomic transitions, which, as discussed above, are absent by construction in the (generalized) Peierls approach, and have to be accounted for separately~\cite{millis_review}.
Therewith, the above derived expression of the Fermi velocity is particularly suited for use in Wannier function setups for
compounds with d or f orbitals, which satisfy the requirement of localized orbitals.

In the appendix, we give explicit expressions for going beyond the generalized Peierls approach, e.g.\ we derive a formula
for including intra-atomic transitions within the current setup of localized orbitals.
In the practical calculations for VO$_2$ within a localized basis, however, we found these terms to be negligible as 
evidenced by the good agreement between the Peierls treatment and experimental findings.

\subsection{Downfolding of Fermi velocity matrix elements -- \\ Upfolding
of the downfolded response}\label{vkdf}

Many-body calculation for realistic systems often work in a downfolded setup.
In other words, after a band-structure has been obtained from \eg an
LDA computation, orbitals that are supposed to be subject to only
minor correlation effects are integrated out and linearized. 
These are typically high energy excitations, and thus
the downfolding procedure is used to construct an effective low-energy problem, which is simpler to be tackled with a many-body approach. The linearization step preserves the Hamiltonian form of the one-particle part of the problem.
Thereby the influence of correlation effects beyond the one-particle \bs\ of these orbitals, and also the possible feedback on the others, are entirely neglected, and the high energy excitation spectrum fixed to the Kohn-Sham eigenvalues within LDA. 
The many-body calculation thus lives in an orbital subspace only, and all other orbital degrees of freedom remain unaffected.
The introduction of a double-counting term, which corrects for correlation effects already taken into account by the LDA, adjusts the center of gravity of the many-body spectrum with respect to the higher energy parts.
See however e.g.\ Ref.~\cite{anisimov:125119} for a way how to include uncorrelated orbitals within the LDA+DMFT cycle. 
For a recent scheme to incorporate also self-energy effects of higher energy orbitals into an effective model see Ref.~\cite{ferdi_down}.

Although in the computation of the Fermi velocities, \eref{matel}, only the ``non-interacting''  Hamiltonian enters, several complications occur, when it comes to deducing of optical properties from downfolded many-body calculations~:
 Not only are transitions from
and to high energy orbitals truncated, but also the optical
transitions within the block of low-energy orbitals acquire wrong
amplitudes. This owes to the fact that, evidently, the computation of transition matrix elements and the downfolding procedure do not commute.
Here one has to distinguish between the effect on the full matrix
element from that on the approximation of the Peierls velocity. 
Indeed, the orbitals of the downfolded system are in general less localized than the ones of the original problem, and the Peierls approximation
therewith is less accurate.
For an instructive discussion on this subject see also \cite{millis_review}.

Here we explain a simple strategy~\cite{tomczak_v2o3_proc} for the computation of the optical conductivity, applicable to many-body electronic structure calculations that were performed using a downfolded one-particle Hamiltonian.
In line with the above remarks, this procedure yields better results than when computing the Fermi velocities directly from the downfolded system.
The procedure is, moreover, not limited to the use of the Peierls approach. 

The central quantity to look at in this respect is the orbital trace of the matrix product of Fermi velocities and momentum-resolved spectral functions in \eref{oc}~:
\begin{equation}
\tr\biggl\{v_\svek{k}A_\svek{k}(\omega\pr)v_\svek{k}A_\svek{k}(\omega\pr+\omega)\biggr\}
\end{equation}
Since the trace is invariant under unitary transformations, the above can be written as
\begin{equation}
\tr\biggl\{U ^\dag_\svek{k} v_\svek{k}U_\svek{k} \widetilde{A}_\svek{k}(\omega\pr)U ^\dag_\svek{k}v_\svek{k}U_\svek{k} \widetilde{A}_\svek{k}(\omega\pr+\omega)\biggr\}
\end{equation}
for arbitrary unitary matrices $U_\svek{k}$.
In the case of a band-structure calculation (\ie a vanishing \se, ${\Sigma}=0$), we can chose these matrices such that they perform the desired downfolding, \ie both, the spectral functions $\widetilde{A}_\svek{k}=U ^\dag_\svek{k} A_\svek{k}U
_\svek{k}$ and the transformed Hamiltonian will acquire a block-diagonal form.
In the following we shall distinguish between the low energy block
``L'' and the high energy block ``H''. An LDA+DMFT calculation will
add local Coulomb interactions only to the former, which will result in a \se\ that lives in the ``L'' sub-block, while the orbitals of the ``H'' block
 will remain unchanged from the many-body (DMFT) calculation.
In other words, since both sub systems are disconnected, the block-diagonality of the spectra is retained throughout the calculation.

The idea is now to compute the dipole matrix elements from the initial full system, and then to apply the same basis transformation that blockdiagonalizes the Hamiltonian also to the velocities.

Clearly the downfolding procedure is not
exact, since it linearizes the impact of the high energy
orbitals. When solving the system with the full, non-downfolded,
Hamiltonian, the matrices that block-diagonalize the full system would not be the same. They would even depend on the frequency $\omega$ due to the dynamical nature of the \se.
Yet, when granting the approximative validity of the downfolding as
such, and assuming the $U_\svek{k}$ to remain unchanged with respect to
the initial band-structure, we can proceed further, and by specifying
\begin{eqnarray}
\widetilde{v}_\svek{k}=U ^\dag_\svek{k} v_\svek{k}U^{\phantom{\dag}}_\svek{k}=\left(
\begin{array}{cc}
V_1 & W\\
W^\dag & V_2\\
\end{array}
\right),\quad
\widetilde{A}_\svek{k}(\omega\pr)=\left(
\begin{array}{cc}
L & 0\\
0 & H\\\end{array}
\right),\nonumber \\ \hbox{and} \qquad
\widetilde{A}_\svek{k}(\omega\pr+\omega)=\left(
\begin{array}{cc}
\bar{L} & 0\\
0 & \bar{H}\\
\end{array}
\right)\label{LH}\qquad\qquad
\end{eqnarray}
where the spectra of the L sector are taken from the many-body calculation.
Then the above trace becomes
\begin{equation}\label{dfvkor}
LV_1\bar{L}V_1 + LW\bar{H}W^\dag +
HV_2\bar{H}V_2+HW^\dag\bar{L}W
\end{equation}
For transitions within the block of only correlated orbitals, $L$ , intervenes the Fermi velocity
matrix $V_1$, which is evaluated as the low-energy block of the unitary transformed matrix element of the {\it full}, \ie non-downfolded system.
The resulting velocity $\widetilde{v}_\svek{k}$ is thus different
from the matrix element that is computed from the downfolded system. When using in particular the Peierls expression
the momentum derivative of the unitary matrices $U_\svek{k}$ lead to
additional terms in the latter case.

Moreover, with the above, a restriction to the low-energy block is not imperative.
We can indeed calculate the complete optical response,
including transitions from, to and within the high energy
block\footnote{We can thus make a distinction between different origins of spectral weight. Yet, we cannot tell apart the different contributions within the $L$ block. While one has the
possibility to suppress selected transitions by setting to zero the
respective Fermi-velocity matrix elements, the different contributions
are in that case not additive.}.
The latter 
is entirely determined
by the
band-structure calculation. When comparing to experiments, this  allows to assess whether
it is only the relative position with respect to the low energy orbitals that needs an ``adjustment'', 
which is connected to the double counting term, correcting for
correlation effects already taken into account by the LDA, or whether correlation effects modify
substantially the overall spectrum of downfolded orbitals. The latter
can be brought about \eg by non-negligible life-times, or shifts that depend on the individual orbital.

We will refer to the above described scheme as ``upfolding'', since the
downfolded orbitals are reintroduced for the sake of accounting for
optical transitions from, into, and between them.

\section{Optical conductivity of vanadium dioxide --  An application of the formalism}

In a recent work~\cite{tomczak_vo2_proc,me_vo2}, we used an
analytical continuation procedure to calculate real-frequency
self-energies from LDA+Cluster-DMFT data\cite{biermann:026404}. This allowed for a better understanding
of the impact of correlation effects, especially for the insulating
phase of \vo2. Indeed, it was revealed that correlation effects enhance the
Peierls effect, while non-local fluctuations preserve the coherence of
one-particle excitations. Moreover we successfully constructed a static, yet orbital-dependent potential that, when added to the LDA Hamiltonian, reproduced the LDA+CDMFT excitation spectrum when artificially neglecting all lifetime effects.
In our picture, M1 \vo2\ is not a genuine Mott insulator, and we
referred to it as a ``many body Peierls'' insulator.

Here, we compute the optical conductivities of both, the metallic and the insulating phase.
A comparison with experimental results 
will allow to further confirm the underlying LDA+CDMFT electronic
structure calculation~\cite{biermann:026404}, and its
interpretation~\cite{tomczak_vo2_proc,me_vo2}. Furthermore,  this will also enable us to analyze the
experimentally measured intensities with a solid theoretical background.

The LDA+DMFT optical conductivities are presented in the Figs.~\ref{fig1} and \ref{fig3} for the metal and the insulator, respectively.
In order to compute the optical conductivity also for high energies, we
employ the upfolding scheme detailed above.
In the many-body Cluster-DMFT calculation~\cite{biermann:026404} all
orbitals other than the vanadium \t2g\ were downfolded. The latter
thus constitute the low energy sector, L, according to \eref{LH}. For the calculation of the Fermi velocities we use a larger
Hamiltonian that comprises for the high energy part, H, in particular the vanadium \egs\ and the
oxygen 2p orbitals, and, moreover, the oxygen 2s\footnote{In the M1
  phase we further include the vanadium 4s and
  4p orbitals.}. 
  We sketchily write s,p,\egs\ in the graphics. 
When indicating that transitions are from s,p,\egs\
into the \t2g\ orbitals, this mainly accounts for transitions from the
occupied O2p into empty \t2g\ orbitals, since, e.g., the \egs\ to \t2g\
transitions are derived only from the little occupied weight of \egs\ character that stems
from hybridizations with occupied orbitals.

When referring to the orientation of the electric field, or the light
polarization, we use the simple monoclinic lattice as
reference\footnote{See e.g. Fig.10 in \cite{eyert_vo2} for the
  first Brillouin zone.}.
 Since for the Peierls Fermi velocity, \eref{dhdk}, we perform the
numerical derivative of the Hamiltonian on a discrete momentum mesh,
not all directions are accessible in a straight forward manner. Yet,
the important polarizations, $E\parallel[001]$ and $E\perp[001]$, are 
capturable. In an experiment, the polarization is varied by choosing
different orientations of the sample, or different substrates, which, in the
case of thin films, favor different growth directions. Herewith, all
orientations that lie within the plane of the surface are probed, when
using unpolarized light. In our
calculations, however, we evaluate the response of a single given
polarization only, without averaging over an ensemble of in-plane directions.

As a comparison to our theoretical curves, we include results from
three experiments that we already mentioned in the beginning.
We will display measurements on single
crystals by Verleur \etal~\cite{PhysRev.172.788}, performed for different
orientations of the sample. Moreover, recently, experiments were
carried out
on different types of thin films. The work of
Okazaki \etal~\cite{PhysRevB.73.165116} used thin films (T$_c\approx290$~K) with $[001]$ orientation, \ie for the electric field $E\perp [001]$. Qazilbash \etal~\cite{qazilbash:205118} on the other hand
used polycrystalline films with preferential $[010]$ orientation (T$_c\approx340$~K).
We now proceed with the presentation of our results for the individual phases.

\subsection{Rutile \vo2\ -- The metal}

\begin{figure}
  \begin{center}
\includegraphics[angle=-90,width=0.48\textwidth]{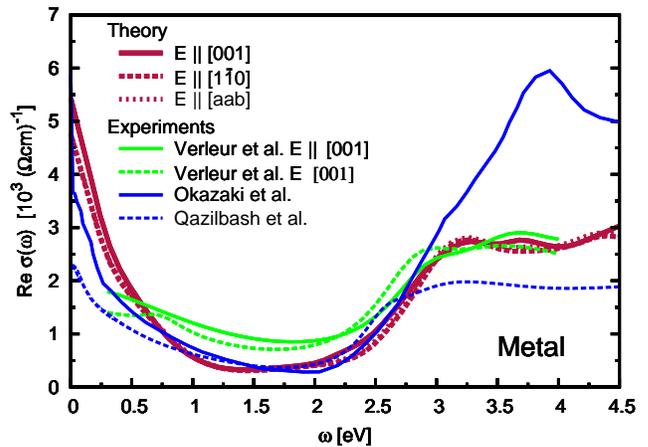}
    \caption{(color online) LDA+CDMFT optical conductivity of the rutile phase of \vo2\ for the indicated polarizations
    ([aab]=[0.85 0.85 0.53]). The velocity matrix elements were calculated using the scheme of \sref{vkdf}. Beyond the \t2g\ orbitals
      this calculation includes in particular the V\egs\ and O2p
      orbitals. Experimental curves from (a)~\cite{PhysRev.172.788}
      (single crystals, orientation as indicated), (b)~\cite{qazilbash:205118}
      (polycrystalline film (T$_c\approx340$~K), preferential orientation $E \perp
      [010]$, T=360~K), and (c)~\cite{PhysRevB.73.165116} (thin
      film (T$_c\approx290$~K), $E \perp [001]$, T=300~K).}
    \label{fig1}
  \end{center}
\end{figure}

In \fref{fig1} we show, along with the mentioned experimental data, the theoretical optical conductivity of rutile
\vo2, which we obtain for the different light polarizations as indicated.
As one can see, already the three experiments yield quite
distinguishable spectra. The differences may
point to a polarization dependence, but one cannot rule out an
influence of the sample type and the means by which multiple
reflections at the sample substrate were treated in case of the thin films.
Indeed, in the case of rutile \vo2, x-ray
experiments~\cite{haverkort:196404} witness a rather isotropic response.
The different measurements on single crystals~\cite{PhysRev.172.788} also
evidence a quite uniform conductivity up to 4~eV.
\begin{figure}
  \begin{center}
\includegraphics[angle=-90,width=0.48\textwidth]{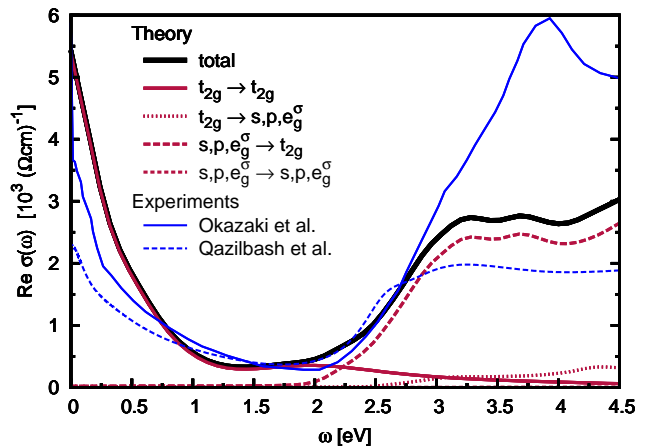}
    \caption{(color online) LDA+CDMFT optical conductivity of rutile \vo2\ for the [001] polarization.
      Shown are the different orbital transitions according to their energy sector, see \eref{dfvkor}. The contributions
      are additive and sum up to the total conductivity. For details see
      \sref{vkdf}. 
Experimental curves, as above from~\cite{PhysRevB.73.165116,qazilbash:205118}.}
    \label{fig2}
  \end{center}
\end{figure}
The polarization dependence of the theoretical conductivity is found
to be rather small, too, which is also in line with our previous
statement~\cite{me_vo2} that the \t2g\ \se\ shows no particular orbital dependence.
Thus, in theory, the metallic Drude-like response is made up from \a1g\ {\it and} \egp\ density near the Fermi surface.

At higher energies, beyond the Drude-like tail, further inter-``band''
intra-\t2g\ transitions occur. Yet, the optical response is rather
structureless up to 2~eV. At this energy, however,
we already expect the onset of oxygen 2p derived
transitions. In order to elucidate the origin of the spectral weight
of this region in greater detail, we plot in \fref{fig2} the optical conductivity
resolved into the different energy sectors, according to \eref{dfvkor}.
Since the O2p and the \egs\ orbitals were part of the downfolded high energy
sector, their position, within our scheme, is frozen to the LDA
result (see \eg the \bs\ in \cite{eyert_vo2}). Therefore transitions from the O2p orbitals into the \t2g\
ones start, as expected, at
around 2~eV.
We remark that the polarization dependence for the oxygen derived
transitions agrees very well with the single crystal
experiments~\cite{PhysRev.172.788} up to 4.5~eV.
Transitions from the \t2g\ orbitals into the \egs\
set in later, at around 2.5~eV, and are rather small in magnitude. 
The O2p to \egs\ transitions appear at the expected energies, but they
are too low to be seen in \fref{fig2}.

Overall, the LDA eigenvalues seem to give a rather good description of
the \egs\ and O2p orbitals, since the agreement with experiment is
reasonably accurate, as was qualitatively noticed already
in previous LDA optics calculations~\cite{0953-8984-19-34-346225}.
When looking at photoemission
results~\cite{PhysRevB.43.7263,koethe:116402}, one remarks that the
on-set of the oxygen 2p is compatible with the LDA, yet, their center of
 gravity is shifted to slightly higher binding energies in the experiment. 
As to the \egs\ orbitals, it is conceivable, when resorting to x-ray
experiments~\cite{PhysRevB.43.7263,koethe:116402} as a reference, that
they appear at a little larger energies and with a
smaller bandwidth than within the LDA.
Of course
both comparisons are somewhat indirect, due to the occurrence of matrix
elements and other effects in the experiments.
Yet, we emphasize that the rather incoherent nature of the \t2g\ weight
in the spectral function~\cite{me_vo2} is far
beyond any band-structure technique, which is why the optical
conductivity in the 2.5 to 4.0~eV region, derived from O2p to \t2g\ transitions, comes out too
large in LDA~\cite{0953-8984-19-34-346225} when comparing to the 
experiment of Ref.~\cite{PhysRev.172.788}, while we find a good agreement for the LDA+CDMFT conductivity.

At this point, we can only speculate on the origin of the shoulder and peak
structure seen in one of the experiments~\cite{qazilbash:205118} at
2.5~eV, and 3.0~eV, see \fref{fig2}. It seems conceivable that it stems from \t2g\ to
O2p transitions, rather than from \egs\ contributions. 
Attributing the humps to distinct O2p to \a1g\ or \egp\ transitions is
cumbersome, mostly due to the structure of the numerous oxygen bands. When looking at the momentum-resolved optical conductivity
(not shown), one realizes that O2p to \egp\ transitions start for most
of the $\vek{k}$-regions at lower energies than transitions into the \a1g.

\subsection{Monoclinic \vo2\ -- The insulator}

We now come to the optical spectra of the monoclinic phase. 
\begin{figure}
  \begin{center}
\includegraphics[angle=-90,width=0.48\textwidth]{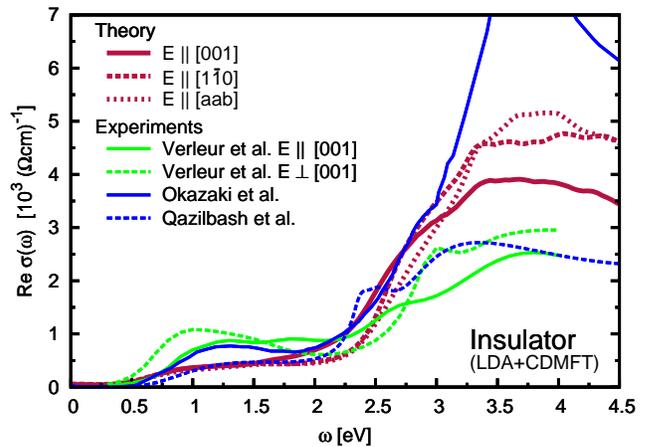}
    \caption{(color online) LDA+CDMFT optical conductivity of the M1 phase of \vo2\ for the indicated polarizations
    ([aab]=[0.84 0.84 0.54]). The velocity matrix elements were calculated using
    the scheme of \sref{vkdf}.
Experimental curves from (a)~\cite{PhysRev.172.788}
      (single crystals, orientation as indicated), (b)~\cite{PhysRevB.73.165116} (thin
      film, $E \perp [001]$, T=280~K), and (c)~\cite{qazilbash:205118}
      (polycrystalline film, preferential orientation $E \perp
      [010]$, T=295~K).}
    \label{fig3}
  \end{center}
\end{figure}
\begin{figure}
  \begin{center}
\includegraphics[angle=-90,width=0.48\textwidth]{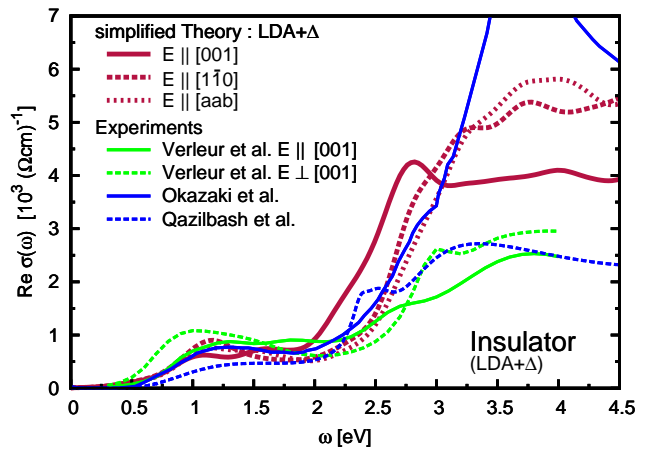}
    \caption{(color online) Optical conductivity of the M1 phase of \vo2\ for the indicated polarizations
    ([aab]=[0.84 0.84 0.54]) when using the effective
    band-structure~\cite{me_vo2} ``LDA+$\Delta$''.
Experimental curves from (a)~\cite{PhysRev.172.788}
      (single crystals, orientation as indicated), (b)~\cite{PhysRevB.73.165116} (thin
      film, $E \perp [001]$, T=280~K), and (c)~\cite{qazilbash:205118}
      (polycrystalline film, preferential orientation $E \perp
      [010]$, T=295~K)}
    \label{fig3a}
  \end{center}
\end{figure}
In \fref{fig3} we show our theoretical LDA+CDMFT results, again, in conjunction with the three
experiments~\cite{PhysRev.172.788,PhysRevB.73.165116,qazilbash:205118}.
As was the case for the metallic phase, the latter yield varying
results. While the optical gap is roughly
0.5~eV in all cases, the higher energy response is markedly different.
Not only the amplitudes, but also the peak positions differ considerably.
Yet, as a matter of fact, in the current case of M1 \vo2, a sizable
polarization dependence is expected from the structural considerations mentioned above.
Indeed our calculation suggests a noticeable anisotropy in the optical
response, which is congruent with the experimental
findings.

Before interpreting the results, however, we find it instructive to also compute the conductivities using a somewhat simpler
approach~: In Ref.~\cite{tomczak_vo2_proc,me_vo2} we deduced from the LDA+CDMFT self-energies
an effective static, yet orbital-dependent, one-particle potential
that reproduced the many-body excitation spectrum, which arises when neglecting all
life-time effects. Therewith all correlation induced
energy shifts are
captured, whereas the coherence of the excitations remains infinite. This is equivalent 
to the use of a scissors operator, albeit one that does not simply widen the gap~\cite{0953-8984-19-34-346225}, but selectively shifts
the one-particle excitations that mediate the dimerization.
We thus replace in a full orbital LDA Hamiltonian the Kohn-Sham
eigenvalues of the \t2g\ orbitals by the ones to which the additional potential was applied, and label the results ``LDA+$\Delta$''.
For further details see~\cite{me_vo2,me_phd}.
Therewith, we do not have to invoke the upfolding scheme of
\sref{vkdf}.

What this theoretical conductivity is missing are the life-time effects
encoded in the imaginary part of the LDA+CDMFT \se. These were found to
be small, yet not entirely negligible~\cite{me_vo2}. 
\fref{fig3} displays our result, again along with the experimental curves.

When looking first at the optical conductivity that results from the effective \bs~\cite{tomczak_vo2_proc,me_vo2}, ``LDA+$\Delta$'', \fref{fig3a}, we
find all experimental polarization tendencies reproduced~: Consistent with Verleur
et al.\cite{PhysRev.172.788}, the $E\parallel[001]$ conductivity is lower than the
$E\perp[001]$ one at energies up to 1.5~eV, after which the c-axis response 
develops a little maximum of spectral weight in both, experiment and theory. At energies of
2.35~eV~\cite{qazilbash:205118} or 3.0~eV~\cite{PhysRev.172.788} the experimental
conductivity with $E\parallel[001]$ components evidences a narrow
peak. In the calculation this is prominently seen at 2.75~eV. When looking at our
effective \bs~\cite{tomczak_vo2_proc,me_vo2}, it seems plausible that these transitions stem from
\a1g\ bonding to anti-bonding orbitals.
The peak is indeed very narrow for an inter-band transition, but in our
picture
this is simply owing to the fact that the \a1g\ anti-bonding excitation
does exhibit an almost dispersionless behavior~\cite{tomczak_vo2_proc,me_vo2}. 
However, already in this frequency region we expect transitions that
involve the oxygen 2p orbitals, as will be detailed below for the
LDA+CDMFT conductivity.

 At still higher energies, the $E\parallel[001]$
response is again lower than for the perpendicular direction in both,
experiment and theory.
The overall congruity with experiments further corroborates the
validity of our effective \bs\ picture for spectral properties and
therewith strengthens our interpretation of the nature of the
insulating phase of \vo2\ as a realization of a ``many-body Peierls'' state~\cite{me_vo2}.

Now we compare this simplified approach with the full LDA+CDMFT conductivities of \fref{fig3}.
We instantly realize that the LDA+CDMFT response for the \t2g\ orbitals is
damped and therewith less structured, which was clearly expected.
The small underestimation of the optical gap is probably owing to the
elevated temperature at which the LDA+CDMFT quantum Monte Carlo
calculation was performed~\cite{biermann:026404}.

\begin{figure}
  \begin{center}
\includegraphics[angle=-90,width=0.48\textwidth]{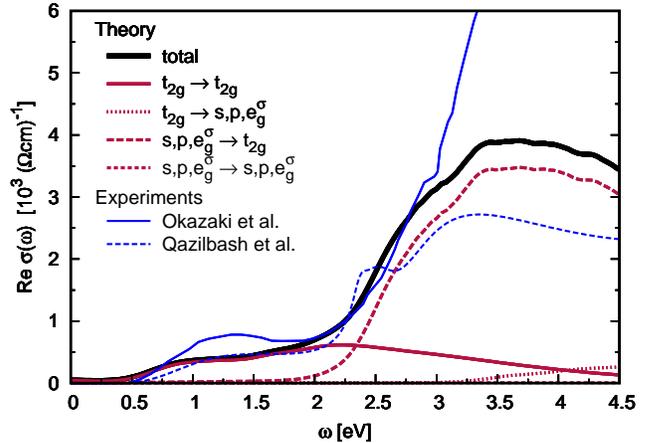}
    \caption{(color online) LDA+CDMFT optical conductivity of M1 \vo2\ for the [001] polarization.
      Shown are the different orbital transitions  according to their energy sector, see \eref{dfvkor}. The contributions
      are additive and sum up to the total conductivity. For details see \sref{vkdf}. 
Experimental curve, as above from~\cite{PhysRevB.73.165116,qazilbash:205118}.}
    \label{fig4}
  \end{center}
\end{figure}

To shed further light on the structure of the response, we resolve in
\fref{fig4} the contributions to the $[001]$ LDA+CDMFT conductivity into their respective energy sectors, according to \sref{vkdf}.
From this we first infer that the slight upturn, seen for this
polarization beyond 1.5~eV in the LDA+CDMFT conductivity is indeed derived from transitions within the
\t2g\ manifold, for oxygen contributions only set in at around 2.0~eV.

Besides, the prominent peak in both, the experimental and LDA+$\Delta$ conductivity with $E\parallel[001]$ polarization
that we attributed above to \a1g--\a1g\ transitions, is largely
suppressed and only faintly discernible as a weak shoulder, when
comparing with the other polarizations. 

As an explanation for this difference between
experiment and the approach of the one-particle potential $\Delta$ on the one hand, and the LDA+CDMFT
result on the other, we forward 
the occurrence of sizable life-time effects in
the LDA+CDMFT electronic structure calculation. Indeed
the \a1g\ spectral weight in the corresponding \t2g\ LDA+CDMFT spectral function is
not sharply defined and extends over more than
2~eV, and is only barely discernible in the total, orbitally traced,
spectrum~\cite{biermann:026404}. When thinking of the conductivity in
simple terms of density-density transitions, it is perfectly
conceivable that the \a1g--\a1g\ response eventuates only in a tail of
spectral weight (as seen in the energy sector resolved conductivity in \fref{fig4}) and not in a well defined peak. Having said this, and
referring to the experiments, one might thus conjecture that these
life-time effects are still overestimated in the
LDA+CDMFT calculation.
Moreover, we stress again that the many-body electronic
structure was computed at high temperatures~\cite{biermann:026404}, which will lead to an overestimation of
the temperature induced part of the broadening~\footnote{We note that only the
LDA+CDMFT scheme makes use of the upfolding procedure of the matrix
elements, whereas the other calculation uses the untransformed Peierls Fermi velocities
of the large Hamiltonian.}.

Finally, we remark that despite all differences in the experimental data,
they reveal (maybe apart from the single crystal for $[001]$ polarization) a common global tendency, namely that, when going from the metal to the insulator, low frequency spectral weight is transfered to higher energies. 
Indeed, for a given polarization, the Drude-like weight that the insulator is lacking at low energies must be recovered, as requires the f-sum rule~\cite{millis_review}. This condition is met at 5.5~eV in one experiment~\cite{qazilbash:205118}, while in the other~\cite{PhysRevB.73.165116}
 an overcompensation appears already at energies beyond 3.5~eV.
Theoretically, when using the LDA+CDMFT conductivity, we find values of 3.73~eV and 4.35~eV, for the $[1\bar{1}0]$ and $[001]$
direction, respectively.

\subsection{Comparison with simpler approaches}

\begin{figure}
  \begin{center}
\includegraphics[angle=-90,width=0.48\textwidth]{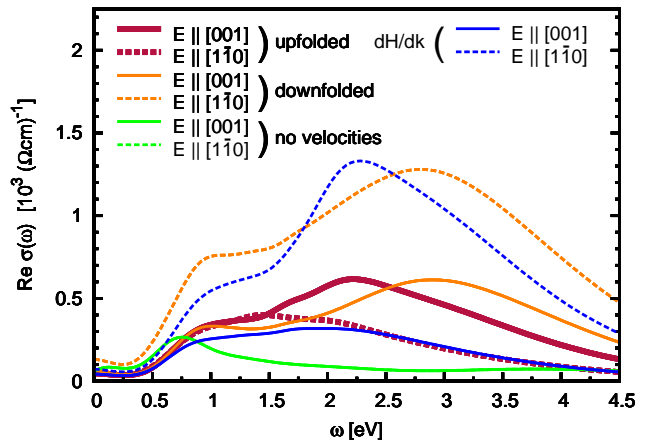}
    \caption{(color online) LDA+CDMFT \t2g\ optical conductivity of the M1
      phase for the indicated polarizations ([aab]=[0.85 0.85 0.53]) and different Fermi
      velocities. The ``upfolded'' curves correspond to our full
      scheme. The ``downfolded'' data computes the Fermi velocities
      from the downfolded Hamiltonian. ``no velocites'' refers
      to a simple convolution of spectral functions without transition amplitudes. The ``dH/dk'' curves correspond to neglecting the multi-atomic generalization in the velocities.}
    \label{fig5}
  \end{center}
\end{figure}

In this section we shall briefly show that all the trouble with the
Fermi velocity is worth the effort.
Therefore we plot in \fref{fig5} for the case of M1 \vo2\ a comparison of our full scheme, which
proved to yield quantitative results, with two simplified calculations.
These differ from the full scheme only in the way how the Fermi
velocities, i.e.\ the transition amplitudes are treated.
We restrict the discussion to the \t2g\ response.

To illustrate the effect of the downfolding of orbitals on the matrix
elements, we have computed the optical conductivity when applying the
generalized Peierls formula on the downfolded Hamiltonian.
As we can see, the resulting curves differ considerably from those using the upfolding
scheme. In particular, the absolute value for some
polarizations is way off with respect to experiment.

It has become a popular approximation to entirely neglect Fermi
velocities
in the computation of optical properties.
Therewith the conductivity is a simple convolution of momentum-resolved
spectral functions. As a consequence inter-band transitions are
omitted, since the Fermi-velocities are simple unit matrices.
Especially in the realistic context this is a severe oversimplification.

Moreover, by construction, there cannot be any orbital dependence in the conductivity,
while, as evidenced from the experiments, this is clearly an important issue for M1 VO$_2$.
Also, the magnitude of intra-band transitions is not properly accounted for.
In fact, the absolute value of the response is not well defined.
In order for this approach to yield a comparable magnitude, we arbitrarily
choose a pre-factor~: $v_{\svek{k}}^{LL\pr}=2r_0\delta_{LL\pr}$, with $r_0$
being the Bohr radius.
As can be inferred from \fref{fig5}, the resulting
peak structure of the optical conductivity is wrong.

Finally, we also compute the optical conductivity when neglecting the
multi-atomic correction term in the velocities, i.e.\ using only the
derivative of the Hamiltonian. As was the case for the velocities of the downfolded
case, only for one polarization does this yield a reasonable result.

\section{Conclusions}
In conclusion we presented a versatile scheme for the calculation of optical properties of correlated materials. Geared at the use with a localized basis set, we devised a realistic extension of the Peierls substitution approach.
Moreover, we developed means to incorporate transitions that involve high energy orbitals that were downfolded in the many-body treatment of the electronic structure.

As an application, we evaluated the optical conductivity of \vo2\ for both, the metallic and the insulating phase. While the metal is characterized by a rather isotropic response, the insulator revealed a noticeable polarization dependence. 
The agreement with experiments is overall satisfying.
The high energy conductivity is reasonably described when using the
LDA band-structure for high lying orbitals. The LDA+CDMFT many-body calculation for the \t2g\
orbitals correctly describes the low-energy behavior. In the rutile
phase it accounts for life-time effects within the \t2g\ orbitals and
therewith also for the damping
of oxygen to \t2g\ transitions with respect to LDA results.
In the insulator, it allowed for a genuine reproduction of the experimental \t2g\ response,
capturing in particular the polarization dependence over a wide energy range.
 The congruity of experiment and theory for the \t2g\ spectral weight can be interpreted as
corroborating the validity of the underlying many-body calculation for the electronic structure along with its interpretation.

\acknowledgments
The authors gratefully acknowledge discussions with 
L. Baldassarre,
N. Bontemps,
A. Georges, 
K. Haule,
H. J. Kim,
G. Kotliar,
A.~I. Lichtenstein, 
R. Lobo,
A.~I. Poteryaev,
M.~M. Qazilbash,
G. Sangiovanni, and
A. Toschi.
This work was supported by Idris, Orsay, under project no. 091393, and the French ANR under project CORRELMAT.


\appendix

\section{Continuum formulation of the transition matrix elements} 

\subsection{Derivation of the generalized Peierls velocity in the continuum}
Starting from the general Fermi velocity, \eref{matel}, which originated from the continuum formulation, we here rederive the generalized Peierls expression as an approximation. The correctional terms contain all intra-atomic transitions, that were completely lacking in the lattice theory, as well as contributions that are owing to the spatial extensions of the wave functions in the solid.

Using $\mathcal{P}_\alpha=-\im m/\hbar\left[\mathcal{R}_\alpha,\op{H}_0\right]$,
the element of \eref{matel} can be written
\begin{eqnarray}
\frac{1}{m}\bra{\svek{k}L\pr}\mathcal{P}_\alpha\ket{\svek{k}L}=
-\frac{\im}{\hbar}\frac{1}{N}\sum_{\svek{R},\svek{R}\pr}e^{\im \svek{k}(\svek{R}-\svek{R}\pr)}\times\qquad\nonumber\\
\int\ddr 
\biggl[
\bra{\vek{R}\pr L\pr}\mathcal{R}_\alpha\ket{\vek{r}}\bra{\vek{r}}\op{H}_0\ket{\vek{R}L} \bigr.\qquad\qquad\nonumber\\
\bigl.
-\bra{\vek{R}\pr L\pr}\op{H}_0\ket{\vek{r}}\bra{\vek{r}}\mathcal{R}_\alpha\ket{\vek{R}L}
\biggr]\quad
\end{eqnarray}
It is important to note, that here the position operator
$\mathcal{R}_\alpha$ is defined in the continuum.
Its effect in the position representation is
$\bra{\vek{r}}\mathcal{R}_\alpha\ket{\vek{R}L}=r_\alpha\chi^{\phantom{*}}_{\svek{R}L}(\vek{r})$.
This is to be contrasted to the discrete lattice version of \eref{Rop}.
Moreover, one has to make a clear distinction between the continuous space variable $\vek{r}$ and the discrete \uc\ label $\vek{R}$.
In the (unphysical) limit of completely localized wave functions, $\braket{\vek{r}}{\vek{R}L}=\chi^{\phantom{*}}_{\svek{R}L}(\vek{r})\sim\delta(\vek{r}-\vek{R}_L)$, 
this distinction is relaxed, and we recover the expression
\eref{vkpeierls} of the Peierls approach, as we shall see.
Like in the lattice case, we split
the atomic positions into $R^\alpha _L=R^\alpha _{\phantom{L}}+\rho_L^\alpha$. Then the above becomes
\begin{eqnarray}
&&\frac{1}{m}\bra{\svek{k}L\pr}\mathcal{P}_\alpha\ket{\svek{k}L}=-\frac{\im}{\hbar}\frac{1}{N}\sum_{\svek{R},\svek{R}\pr}e^{\im \svek{k}(\svek{R}-\svek{R}\pr)}
\times\qquad\qquad\nonumber\\ 
&&\biggl\{(R_\alpha\pr-R_\alpha+\rho^\alpha_{L\pr}-\rho^\alpha_L)
\bra{\vek{R}\pr L\pr}\op{H}_0\ket{\vek{R}L}\biggr. +\nonumber\\
&&\int\ddr r_\alpha 
\bigl[ 
\braket{\vek{R}\pr L\pr}{\vek{r}+\vek{R}\pr+\mop{\rho}_{L\pr}}\bra{\vek{r}+\vek{R}\pr+\mop{\rho}_{L\pr}}\op{H}_0\ket{\vek{R}L} \bigr. \nonumber\\
&&\qquad\biggl.\bigl.-
\bra{\vek{R}\pr L\pr}\op{H}_0\ket{\vek{r}+\vek{R}+\mop{\rho}_L}\braket{\vek{r}+\vek{R}+\mop{\rho}_L}{\vek{R}L}
\bigr]\biggr\}\label{intermed2}
\end{eqnarray}
where we have chosen to condense everything into two different terms.
The first one obviously is
\begin{equation}
\label{dhdk}
\frac{1}{\hbar}\biggl(\frac{\partial}{\partial k_\alpha} \bra{\vek{k}L\pr}\op{H}_0\ket{\vek{k}L}-\im(\rho^\alpha_{L\pr}-\rho^\alpha_L)\bra{\vek{k}L\pr}\op{H}_0\ket{\vek{k}L}\biggr)
\end{equation}
which is exactly the generalized Peierls expression, \eref{vkpeierls}. 
The merit of the Peierls approximation, in particular in realistic calculations, is its
apparent simplicity. Indeed no matrix elements other than the Hamiltonian need
to be calculated. The latter is  a quantity that is anyhow required
for a many-body calculation, and one thus has only to perform the
directional momentum derivative. \footnote{We perform this derivative by
  using the four-point formula~:
  $f\pr(x_i)dx\approx\frac{1}{12}\left(f(x_{i-2})-8f(x_{i-1})+f(x_{i+1})-f(x_{i+2}) \right)$.}
From the discussion of the Peierls substitution above, it is clear that the second term in \eref{intermed2}
accounts on the one hand for all atomic transitions
($\vek{R}=\vek{R}\pr$ and $\gamma\pr=\gamma$), yet it also contains contributions that arise
from the fact that we started from a continuum
formulation. In other words, the spatial extensions of the wave functions lead to
inter-atomic, $\gamma\pr\ne\gamma$, corrections, owing to their finite overlap.
Yet, a direct evaluation of these terms is an intricate undertaking, since it involves the calculation of many integrals. Therefore, it is a valid question whether the generalization of the Peierls approach, as such, already gives a reasonable approximation, without considering the terms beyond it, and if so, under which circumstances.
Though the regrouping of terms into the Peierls expression and the rest was guided from the lattice considerations, it might still seem somewhat arbitrary. The next section however reveals that the intuition of an increased validity of the Peierls approach with a better localization of the involved orbitals is actually warranted.  

\subsection{The Peierls substitution as the localized limit}

In the following, we will make consecutive approximations regarding
the extension of the orbitals, which lead, step by step, to more simplified correction terms to the Peierls expression \eref{dhdk}, which one might endeavor to take into account in an actual computation. Moreover, 
these approximative steps will rationalize the identification of the Peierls term as the leading contribution to the Fermi velocities in the considered setup.

By assuming well localized orbitals, we thus proceed to cut down the  expression in question to the predominant terms, which will be given by the integrals that involve wave functions that have a large overlap \footnote{Although the matrix element is not a mere overlap, and it is actually conceivable that in some cases ``non-local'' terms are important, this constitutes an improvement to the approximation that we are to consider.}.
Indeed, we show that in the limit of perfect localization (or
equivalently in the limit of large atomic separation) the only
surviving transition elements are given by the intra-atomic
contributions, that were missing in the Peierls formulation. 
Using $\braket{\vek{r}+\vek{R}}{\vek{R}L}=\chi^{\phantom{*}}_{\svek{R}L}(\vek{r}+\vek{R})=\chi^{\phantom{*}}_{\svek{0}L}(\vek{r})$, the terms beyond the Peierls ones can be put into the form
\begin{eqnarray}
&&-\frac{\im}{\hbar}\frac{1}{N}\sum_{\svek{R},\svek{R}\pr}e^{\im \svek{k}(\svek{R}-\svek{R}\pr)}\int \ddr r_\alpha \sum_{\mop{\Lambda}, L\ppr}\\
&&\biggl[
\chi_{\svek{0} L\pr}^*(\vek{r}+\mop{\rho}_{L\pr})\chi^{\phantom{*}}_{\svek{0} L\ppr}(\vek{r}+\vek{R}\pr-\mop{\Lambda}+\mop{\rho}_{L\pr}) \bra{\mop{\Lambda}
  L\ppr}\op{H}_0\ket{\vek{R}L} \biggr. \nonumber\\
&&-\biggl. \chi_{\svek{0} L\ppr}^*(\vek{r}+\vek{R}-\mop{\Lambda}+\mop{\rho}_L)\chi^{\phantom{*}}_{\svek{0}L}(\vek{r}+\mop{\rho}_L)
\bra{\vek{R}\pr L\pr}\op{H}_0\ket{\mop{\Lambda} L\ppr}
\biggr]\nonumber
\end{eqnarray}
In this formula the origins of all intervening wave functions lie within the same
\uc, labeled ``$\vek{0}$''~\footnote{Depending on the structure, however, atoms in neighboring cells might be in closer a vicinity than other atoms in the same cell.
}%
. In a first step, the assumed localization of the involved orbitals makes it
reasonable to identify important terms in the sum as those, where the arguments of the wave functions also lie within the same \uc, \ie 
$\mop{\Lambda}=\vek{R}\pr$ in the first term and $\mop{\Lambda}=\vek{R}$ in
the second one.
We note that within this approximation, only the Hamiltonian element depends on the
\uc\ labels $\vek{R}$ and $\vek{R}\pr$, and we can thus directly perform the Fourier transformation, yielding
\begin{eqnarray}\label{intermed3}
&&-\frac{\im}{\hbar}\int \ddr r_\alpha \sum_{L\ppr}\\
&&\quad\biggl[
\chi_{\svek{0} L\pr}^*(\vek{r}+\mop{\rho}_{L\pr})\chi^{\phantom{*}}_{\svek{0} L\ppr}(\vek{r}+\mop{\rho}_{L\pr}) \bra{\vek{k}L\ppr}\op{H}_0\ket{\vek{k}L} \biggr. \nonumber\\
&&\qquad-\biggl. \chi_{\svek{0} L\ppr}^*(\vek{r}+\mop{\rho}_L)\chi^{\phantom{*}}_{\svek{0}L}(\vek{r}+\mop{\rho}_L)
\bra{\vek{k} L\pr}\op{H}_0\ket{\vek{k} L\ppr}
\biggr]\nonumber
\end{eqnarray}
This means that the entire momentum dependence,
in this approximation, is carried by the Hamiltonian.
The complexity of the occurring matrix elements of the position operator has been considerably reduced.
In the {\it one-atomic case}, \ie $\gamma=\gamma\pr=\gamma\ppr$, and when using the short-hand notation
$\op{R}_{\alpha,0}^{LL\pr}=\bra{\vek{0}L}\mathcal{R}_\alpha\ket{\vek{0}L\pr}$, 
$\op{H}_{0}^{LL\pr}(\vek{k})=\bra{\vek{k}L}\op{H}_0\ket{\vek{k}L\pr}$
we simply have
\begin{eqnarray}\label{effatomp}
        -\frac{\im}{\hbar}\com{\op{R}_{\alpha,0}
        ,\op{H}_{0}(\vek{k})}_{L\pr L}
\end{eqnarray}
This is reminiscent of the relation $1/m\mathcal{P}=-\im/\hbar\com{\mathcal{R},\op{H}_0}$, which we used in the beginning.
Here however intervene on-site matrix elements rather than the full position operator. 
Indeed these elements,
$\op{R}_{\alpha,0}^{LL\pr}$,
are well known in atomic physics~: They give the usual amplitudes for
atomic dipolar transitions~: The angular part of the integral will
produce the corresponding dipole selection rules ($\Delta l=\pm 1, \Delta m=0,\pm 1$) via
Clebsch-Gordon coefficients (see e.g. \cite{tannoudji}), when, as we have assumed, the wave functions
have a well-defined angular momentum $(l,m)$.
Contrary to the atomic case, however, the Hamiltonian is momentum
dependent, owing to the fact that, though regarding atomic
transitions, the ``atom'' is here embedded in a solid. 
Also, the above term reminds the form of the multi-atomic correction term in \eref{dhdk}, only that there occurred fixed atomic positions $\mop{\rho}_\gamma$, which commute with
the Hamiltonian, which is why in \eref{dhdk} only the non-local terms $\gamma\ne\gamma\pr$ appear.  
 
Coming back to the multi-atomic case, we have to make a further approximation in order to obtain an expression containing atomic transitions only. Yet, the shifts in the wave functions of \eref{intermed3} can be treated analogous to the \uc\ coordinates~: Indeed $\chi^{\phantom{*}}_{\svek{0}L}(\vek{r}+\mop{\rho}_L)$ is centered around the
position of atom $\gamma$. When, for the sake of clarity, we rename $\widetilde{\chi}^{\phantom{*}}_{\svek{0}L}(\vek{r})=\chi^{\phantom{*}}_{\svek{0}L}(\vek{r}+\mop{\rho}_L)$
we find
\begin{eqnarray}
&&-\frac{\im}{\hbar}\int \ddr r_\alpha \sum_{L\ppr}\\
&&\quad\biggl[
\widetilde{\chi}_{\svek{0} L\pr}^*(\vek{r})\widetilde{\chi}^{\phantom{*}}_{\svek{0} L\ppr}(\vek{r}+\mop{\rho}_{L\pr}-\mop{\rho}_{L\ppr}) \bra{\vek{k}L\ppr}\op{H}_0\ket{\vek{k}L} \biggr. \nonumber\\
&&\qquad-\biggl. \widetilde{\chi}_{\svek{0} L\ppr}^*(\vek{r}+\mop{\rho}_L-\mop{\rho}_{L\ppr})\widetilde{\chi}^{\phantom{*}}_{\svek{0}L}(\vek{r})
\bra{\vek{k} L\pr}\op{H}_0\ket{\vek{k} L\ppr}
\biggr]\nonumber
\end{eqnarray}
From this expression it is plausible, that for localized orbitals atomic transitions
($\gamma\ppr=\gamma\pr$ and $\gamma\ppr=\gamma$, respectively) are in fact
predominant. When restraining ourselves to these cases, we thus drop
entirely the corrections to hopping processes that stem from the
finite extensions of the wave-functions and end up with
\begin{eqnarray}
&&-\frac{\im}{\hbar}\int \ddr r_\alpha \left[ \sum_{L\ppr}^{\gamma\ppr=\gamma\pr}
\widetilde{\chi}_{\svek{0} L\pr}^*(\vek{r})\widetilde{\chi}^{\phantom{*}}_{\svek{0} L\ppr}(\vek{r}) \bra{\vek{k}L\ppr}\op{H}_0\ket{\vek{k}L}\right. \nonumber\\
&&\left.
-  \sum_{L\ppr}^{\gamma\ppr=\gamma} \widetilde{\chi}_{\svek{0} L\ppr}^*(\vek{r})\widetilde{\chi}^{\phantom{*}}_{\svek{0}L}(\vek{r})
\bra{\vek{k} L\pr}\op{H}_0\ket{\vek{k} L\ppr}
\right]\nonumber\\
&=&-\frac{\im}{\hbar} \left[
  \sum_{L\ppr}^{\gamma\ppr=\gamma\pr}\bra{\widetilde{\vek{0}L\pr}}\mathcal{R}_\alpha\ket{\widetilde{\vek{0}L\ppr}} \bra{\vek{k}L\ppr}\op{H}_0\ket{\vek{k}L}\right.\nonumber\\
  &&\left.-\sum_{L\ppr}^{\gamma\ppr=\gamma}\bra{\widetilde{\vek{0}L\ppr}}\mathcal{R}_\alpha\ket{\widetilde{\vek{0}L}} \bra{\vek{k}L\pr}\op{H}_0\ket{\vek{k}L\ppr}\right]\label{intermed1}
\end{eqnarray}
Here, only the terms in which the Hamiltonian element is diagonal in the atomic index $\gamma$ can be written in the form of a commutator, as was the case in the one-atomic case in \eref{effatomp}. In total,
the above term contains
intra-atomic transitions only. These were completely missing in the
Peierls approach, as explained above. Under the assumptions on the localization of the involved orbitals, 
the Peierls term, \eref{dhdk}, thus turns out to be the most important contribution to the Fermi velocity. 
Intuitively, this approach is thus particularly suited for systems in which \eg 3d or 4f orbitals play an important
role, since these verify the request of a high degree of localization.

Another point, worth noticing, is the fact, that, when using the
Peierls approximation, the result of the conductivity is actually
basis dependent. It is the momentum derivative of the Hamiltonian,
which constitutes the first term in \eref{dhdk}, that transforms
evidently differently than the Hamiltonian itself. Obviously this is an artifact of the approximations from
which the Peierls expression eventuated. Since, however, the term
corresponds to the  limit of perfect localization, it is expected to still yield reasonable results for orbitals that are short-ranged. We will come back to this in the next paragraph, in the context of Fermi velocities for downfolded Hamiltonians.

Improvements to the above approximations are obvious: One could e.g.\
take into account elements containing nearest-neighbor wave functions
within the same \uc, or 
even account for wave functions centered in different \uc s.
An evaluation of these terms in principle allows for a more quantitative assessment of the quality of the Peierls term.
However, the matrix elements that one needs to
evaluate are numerous and more complex since they explicitly involve
 various wave functions.
We stress again, that these terms are inter-atomic corrections
to the Peierls term, while the intra-atomic contributions are
absent in the Peierls formalism by construction, and given by \eref{intermed1}.


\end{document}